# Protecting the Future Grid: An Electric Vehicle Robust Mitigation Scheme Against Load Altering Attacks on Power Grids

Mohammad Ali Sayed [a, *], Mohsen Ghafouri [a], Ribal Atallah [b], Mourad Debbabi [a], Chadi Assi [a]

a: Concordia Institute for Information Systems Engineering, Concordia University, Montreal, Quebec, Canada

b: Hydro-Quebec Research Institute, Montreal, Quebec, Canada

*Corresponding Author. Email: mohammadali.sayedd@gmail.com

*Abstract*— Due to the growing threat of climate change, the world's governments have been encouraging the adoption of Electric Vehicles (EVs). As a result, EV numbers have been growing exponentially which will introduce a large EV charging load into the power grid. On this basis, we present a scheme to utilize EVs as a defense mechanism to mitigate Load-Altering (LA) attacks against the grid. The developed scheme relies on robust control theory and Linear Matrix Inequalities (LMIs). Our EV-based defense mechanism is formulated as a feedback controller synthesized using H-2 and H-∞ control techniques to eliminate the impact of unknown LA attacks. The controller synthesis considers the grid topology and the uncertainties of the EV connection to the grid. To demonstrate the effectiveness of the proposed mitigation scheme, it is tested against three types of LA attacks on the New England 39-bus grid. We test our mitigation scheme against 800 MW static, switching, and dynamic attacks in the presence of multiple sources of uncertainty that can affect the EV load during deployment. The results demonstrate how the grid remains stable under the LA attacks that would otherwise lead to serious instabilities.

## 1. INTRODUCTION

Humanity's increasing reliance on electricity has transformed the course of society's development over the past couple of centuries [1]. The power grid has become the center of any advanced society and its security and stability are at the center of any country's national security. To this end, smart technologies have been introduced to support reliable grid operation transforming it into a smart grid [2] [3]. The smart grid, however, became an interconnected system of physical and cyber components leaving it open to attacks initiated through the cyberinfrastructure that can have detrimental impacts on its stability and security [4].

One such attack is the False Data Injection (FDI) attack [5] in which attackers tamper with the grid's measurements to manipulate the state estimation and cause operators to take actions that might damage the grid. Stealthy FDI attacks also remain hidden from the Bad Data Detection (BDD) mechanism employed by utilities, even when attackers have incomplete topology information [5]. To this end, multiple attempts have been made to secure the communication layer of the grid [5] [6]. Yet Load-Altering (LA) attacks against the grid demand side, rather than state estimation, can only be seen through their impact [7] bypassing the BDD.

LA attacks can be broadly classified into 3 subfamilies which are static attacks [7] [8] switching attacks [9] [10], and dynamic attacks [11] [12]. These LA attacks are stealthier and stronger than attacks targeting the grid's cyber layer alone as demonstrated below. The authors of [7] and [8] demonstrated how static attacks can be initiated by manipulating smart home high-wattage Internet of Things (IoT) devices to cause line tripping and load shedding while remaining unobservable to the utility. The switching attacks proposed in [8] manipulate distribution feeders to cause a disturbance that led to generator tripping while mimicking natural phenomena making them hard to be detected by the utility. Finally, the dynamic attacks in [11] [12] achieved grid instability and blackouts by targeting smart loads which cannot be directly monitored by the utility.

Most studies related to cyber intrusions into the power grid focus on attack detection on the cyber layer with little focus on mitigation. The authors of [10], for instance, utilized Neural Networks (NN) to detect switching attacks and achieved a 70% detection accuracy after examining 20s of charging requests data. Although the NN in [13] achieved near-perfect accuracy, it still requires 5s after the attack is initiated to classify it correctly by which time certain attacks would have already damaged the grid. The authors of [14] proposed an accurate detection algorithm based on extremely randomized trees to detect FDI attacks but ignored attack mitigation. On the other hand, most protection mechanisms found in the literature target very specific phenomena and disregard others. The authors of [8], for example, consider that the current N-1 contingency criterion is enough to overcome the impact of static attacks. The authors, however, propose a variation of the attack that causes load shedding even in the presence of N-1 contingency. The switching attack mitigation scheme in [10] uses a wide area controller to mitigate switching attacks with a frequency below 2 Hz making this scheme less effective against higher frequency attacks and dynamic attacks. An optimal output feedback controller was used in [15] to eliminate interarea oscillation following a contingency on the grid but not persistent attacks.

These examples, however, should not overweigh the advantages introduced by the smart technologies incorporated into the smart grid. One such technology is the EV and its charging infrastructure. Faced with the presented reality of the grid's vulnerability to LA attacks, we intend to create an LA attack mitigation scheme that takes advantage of the EVs' unique properties that are ideal for such a purpose. EVs can support the power grid by acting as distributed battery storage as well as distributed generators owing to the Vehicle-to-Grid (V2G) power flow capability in new EV Charging Stations (EVCSs). These EV loads are also spread throughout the power grid such that their distribution covers all load buses. This widespread distribution makes EVs optimal for usage in a wide area controller. This distribution also means that the EVs are collocated with the other system loads that will be used by adversaries for LA attacks. This colocation gives EVs an edge since the disturbances can be efficiently eliminated at their source bus with minimal propagation to the rest of the grid. Finally, EVs have another advantage over generators when used to mitigate fast switching attacks, which is the speed at which EVs can change their load. Turbine generators are rotating machines, and their reaction times are determined by their size, type, weight, and control mechanisms and are usually in the order of several seconds [16] [17]. On the other hand, EVCSs are based on bidirectional power converters [18] that can change their charging rate and toggle between on, off, and V2G instantly [17] in the order of 1ms. This fast reaction time is needed to react to LA attacks especially those initiated from converter-based IoT loads.

Previous studies that have considered using EV loads to support the power grid fall short of achieving the mitigation capabilities suggested in this study. Most of these studies, some of which are discussed in Section 2, focus on using EVs to support the power grid during its steady-state operation. Other studies use EVs passively for load balancing [19].

Based on the above discussion, in this paper, we create a robust wide-area controller based on mixed H-2/∞ controller synthesis that utilizes the EVs as its control inputs to mitigate the impact of persistent static, switching, and dynamic attacks even when they are sustained for long durations. We follow a detailed methodology to evaluate the performance of the controller and examine its performance in comparison to H-2 and H-∞ controllers. The performance of this family of controllers makes them an ideal starting point for our mitigation scheme. The Linear Matrix Inequalities (LMIs) of the controllers are modified to fit our system and incorporate the uncertainties of the EV connection to the grid resulting in the formulation of a family of robust controllers. The robust controller formulation is meant to overcome the deployment obstacles that can cause uncertainties in the control signal sent from the utility to the participating EVs. To the best of our knowledge, this is the first work to consider using EV charging loads in a scheme meant to mitigate the 3 known types of LA attacks. The contributions of this paper can be summarized as:

- We are the first to propose a robust mixed sensitivity wide-area controller that utilizes EV active and reactive charging load to stabilize the power grid during the 3 known types of LA attacks. Our control mechanism successfully eliminates the impact of persisting attacks without the need for a detection mechanism. Our mixed controller eliminates over 99.6% of the attack impact and returns the frequency to its normal operating range instantaneously.

- We design our robust feedback controller to account for uncertainties introduced by real-life deployment obstacles. We mathematically model this uncertainty and incorporate it into our controller synthesis. We are the first to study the stability of wide-area controllers under these uncertainties in the context of smart grid cyber-physical security.

- We demonstrate the effectiveness of our proposed EV-based mitigation mechanism through extensive time-domain simulations. These simulations show how the devastating impact of the 3 known types of LA attacks is eliminated completely and instantaneously while having a negligible impact on the range of the participating EVs as well as negligible cost.

The rest of the paper is organized as follows. Section 2 briefly presents the system preliminaries and the related studies. Section 3 discusses the grid modeling and the mathematical formulation of the controllers. Section 4 discusses the case studies and Section 5 examines the scheme's stability and the effects of uncertainties. Finally, Section 6 concludes the paper.

## 2. PRELIMINARIES AND RELATED STUDIES

In this section, we present the EV numbers and take a brief look at the EV technology we utilize in our mitigation scheme. We then briefly discuss the LA attacker models that require such a mitigation scheme to be present. Finally, we present the related studies in the field of EVs and power grid protection.

### 2.1. EV NUMBERS AND TECHNOLOGY

The world's governments have been encouraging the adoption of EVs to reduce the emissions of the transportation sector. As such, we are witnessing exponential growth in EV numbers on the road. This exponential trend is demonstrated by the record EV sales reaching 3.2 million EVs in 2020 [20] and 6.6 million in 2021 [20]. The trend has continued with 10.5 million EV sales in 2022 and an anticipated 14 million EV sales in 2023 according to the IEA [21].

To support this rapid deployment, EVCS manufacturers have been introducing faster and cheaper EVCSs. While Level 2 EVCSs had a rate of 7.2 kW a few years ago, it has now increased to 11 kW [18] and 19 kW. Level 3 fast EVCSs have rates between 40kW and 360 kW [18]. While all Level 3 EVCSs are DC chargers, Level 2 can be AC or DC. Nonetheless, DC chargers are becoming more common owing to their bidirectional inverter/converter circuits [18] that allow higher charging rates and support the V2G functionality. Furthermore, the current EV infrastructure for public EVCSs provides us with the communication and control mechanisms needed for our mitigation scheme. Public EVCSs are connected to a Cloud Management System (CMS) [18] [22] [23]. This CMS utilizes the Open Charge Point Protocol (OCPP) to communicate, monitor, and control all functionalities of the EVCSs in real-time. By utilizing this underlying infrastructure, our control mechanism removes the need for the addition of any new control software or hardware. These EVCSs are connected to the internet through onboard 5 routers. 5G networks are intended to function in areas of high user density and achieve a speed of 10 Gigabits per second (Gbps) and a latency of 1 millisecond [24]. This is the fastest technology currently available to connect widely disbursed users and achieve reliable and fast communication, making it ideal for our fast-reacting EV-based mitigation scheme. Moreover, OCPP specifies the Network Time Protocol (NTP) as the main protocol used to ensure synchronization of the EVCS clocks [25] [26]. NTP ensures the synchronization of the EVCSs' clocks with a guaranteed accuracy of 10ms over the public Internet [27] and 1ms over a local area network [27].

## 2.2. LA Attack Types and Attacker Model

As mentioned above, LA attacks manipulate actual power consumption to harm the grid [7]. Soltan et al. [7] proposed a family of large-scale static attacks, sudden spikes in load, against the grid using a botnet of compromised high-wattage IoT devices. These attacks only need high-level geographical information on the IoT devices' distribution in the grid. Their attacks cause frequency instability, increase operation costs, or cause line tripping. The authors found that compromising a small fraction of the available water heaters in a grid is sufficient to disrupt its operation. Moreover, the authors of [8] presented a multi-step static attack based on the grid's transient conditions. This variation requires the attacker to have the ability to monitor the grid's transient behavior to launch the attack steps accordingly.

Switching attacks are another form of LA attack that can be launched to excite certain unstable modes present in the grid, e.g., inter-area oscillation [9] [10]. No topology information is required for this attack. However, during the reconnaissance phase, the attacker introduces a chirp signal into the grid using the compromised loads. This signal allows the attacker to monitor the grid's response to different attack frequencies and calculate the impulse response of the system. From this response, the attacker can now determine the specific frequency that would excite the existing unstable mode. The attacker now switches the compromised loads on/off at this specific frequency. The inter-area oscillation mode was excited by a switching attack in [9] and [10]. However, the largest impact of the switching attack can only be achieved when it is done at a frequency of an existing unstable mode.

The third LA attack is the dynamic attack described in [11] and [12]. During the reconnaissance phase, the attacker gathers information about the power grid's topology and parameters to build a state-space model of the system [11]. This state-space is then used to craft the attack as a feedback controller that manipulates the magnitude and oscillation frequency of the compromised load to shift the grid's eigenvalues to unstable operation regions. In [12] the feedback gain was calculated using LMIs and caused the generators to oscillate against each other and the frequency to deviate beyond the 2.5% limit tripping the generators. The dynamic attack load is tailored to the instantaneous changes in the grid's response. The authors of [12] also demonstrated that the success of the attack does not require using 100% accurate grid parameters.

The above examples stress the necessity for a fast-reacting protection scheme against LA attacks. This is especially true since the world is witnessing an increasing ability of attackers to manipulate high-wattage IoT devices as demonstrated by studies performed in partnership with multiple utilities [8] and cyber security companies [28].

## 2.3. Related Studies

Multiple studies have considered using EV charging to support the power grid in its steady-state operation. The work in [29] for example discusses the scheduling of EV charging at non-unity power factor to inject or draw reactive power into the grid. The reactive power flow is then included in the optimization of the EV charging schedule [29]. This strategy reduced the overall cost of EV integration into the distribution grid and improved voltage steady-state stability [29]. A similar study was performed in [30] in which the bidirectional EV charging is optimized to perform peak shaving for the grid during peak demand times. EVs were also considered in [31] as a virtual distributed storage system to mitigate the intermittency of wind generation. The EV charging is optimized to store excess wind generation and then inject it back into the grid at times when wind generation was lower than expected. Another EV usage to support the power grid was suggested in [32] where the authors suggested a local control scheme for 3-phase EV chargers coupled with photovoltaic inverters to balance 3-phase distribution grids. The EVCSs would draw the power from the lightly loaded phases while the inverters inject power into the highly loaded phases. The authors were able to improve 3 phase balance and reduced the power losses by 28%. Other studies have considered EV-based frequency regulation mechanisms against disturbances caused by renewable energy intermittency [33] [34] [35]. As a result, the control mechanisms in [33] [34] are designed to handle frequency deviations below 0.07Hz. Furthermore, the work performed in [35] is designed to deal with frequency fluctuations below 0.06Hz with the occasional sudden spike or drop in renewable energy output power. As such this study is optimized to deal with a single sudden spike or drop in generation and not persistent LA attacks. On the other hand, in our work, we will utilize the EVs to support the small-signal stability of the transmission grid against persistent LA attacks by treating these attacks as persisting disturbances to be attenuated through the action of our EV-based mitigation scheme.

## 3. SYSTEM MODELING AND MITIGATION METHODOLOGY

In the following section, we discuss the defender model and the synthesis of our proposed EV-based controller to be used as our mitigation scheme against LA attacks targeting power grids. In this section, we explain the utility's state-space model of the grid. We then present our observer design that is needed to recover and incorporate the power grid's states into the state-space model. We then move on to discuss our EV-based mitigation controller selection and walk through the synthesis of the controller to fit our scenario and achieve the desired results. Finally, we address the looming issue of the impact on EV users under such a scheme.

### 3.1. POWER SYSTEM REPRESENTATION AND DEFENDER MODEL

As a utility, the defender is assumed to have all knowledge of the grid's parameters and is thus able to represent its behavior with extremely high accuracy. In our study herein, the power system's dynamic behavior is considered mostly dependent on the generators and their control systems. For the modeling of the generators and control systems, we use models which are widely accepted in similar studies [36], i.e., (i) round rotor synchronous machine with order 6, (ii) generator exciter Model IEEE T1, (iii) single mass IEEE G2 steam turbine prime mover, and (iv) Power System Stabilizer (PSS) based on IEEE Std 421.2. We assume that the defender has perfect knowledge of these parameters as well as the line and load parameters to represent the grid. Finally, the grid is linearized into a state-space model with the active and reactive power of the aggregate EV-defender loads as the inputs and the generator frequencies being the outputs. Additionally, the unknown LA attack is represented by a disturbance to the grid. This representation is expressed in (1) where $x, y, u,$ and $\omega$ represent the vectors of system states, outputs, inputs, and disturbances, respectively. Additionally, $\Delta P_{EV_n}$, $\Delta Q_{EV_n}$, and $f_{Gen_m}$ represent the change in active and reactive power of the aggregate EV load at bus n and the frequency of generator m respectively. A, B, C, and D are the state-space matrices that represent the power grid and its dynamic behavior. $B_d$ and $D_d$ represent the impact of the LA attacks on the grid's states and outputs respectively.

$$\dot{x} = Ax + Bu + B_d\omega \quad (1a)$$
$$y = Cx + Du + D_d\omega \quad (1b)$$
$$\text{such that} \quad y = (f_{Gen_1} \ f_{Gen_2} \ \cdots \ f_{Gen_m}) \quad (1c)$$
$$\text{and} \quad u = \Delta PQ_{EV} = (\Delta P_{EV_1} \ \Delta Q_{EV_1} \ \cdots \ \Delta P_{EV_n} \ \Delta Q_{EV_n}) \quad (1d)$$

The authors of [37], [38], and [39] discuss how the power grid can be linearized to maintain all its behavioral properties while the work presented in [40] and [41] discusses how power grids are linearized for the sake of designing their control techniques. Finally, the details of modeling the power grid as a state-space are presented in [42] and [12].

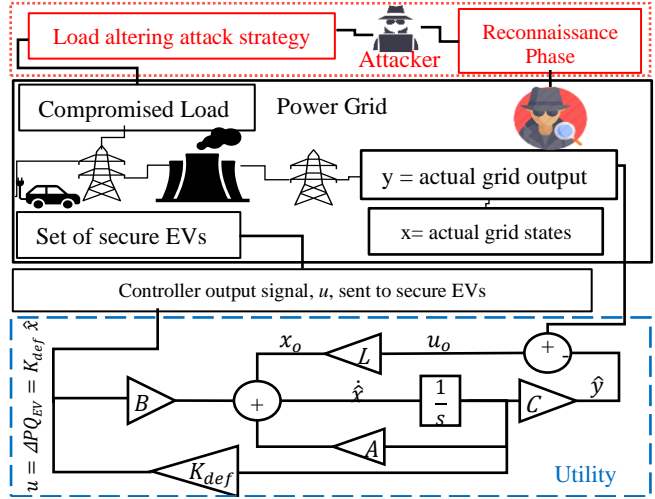
Fig. 1. Defender's state-space model and interaction with the attacked grid

Fig. 1 represents our system model including the LA attack against the grid and the defender's mitigation scheme we are proposing. The attacker in the dotted box above the grid relies on the LA attack models discussed above. The state-space representation in the dashed box, below the grid, is constructed by the utility as part of the EV-based controller scheme we are proposing. This state-space representation is then used to calculate the EV gain matrix $K_{def}$ used to determine the required EV active and reactive power and capture the behavior of the input $u = \Delta PQ_{EV}$ for the system to eliminate the impact of the attack/disturbance. The relation between the disturbance and the generator frequency can be expressed as $y = T\omega$, where $T$ is a transfer function written in terms of the state-space matrices. The behavior of $u = \Delta PQ_{EV}$, is then captured and replicated on the set of secure EVs located throughout the grid. The mitigation controller is designed as a full-state feedback controller, i.e., $u = K_{def}x$, where $K_{def}$ is the gain matrix that is optimized to eliminate the disturbances caused by the attack. The state-space matrices are also used to calculate the observer gain matrix $L$ needed by the utility to recover the grid states, $x$. Since not all the states are measurable, the utility obtains an estimate, $\hat{x}$, by multiplying the grid's outputs, y, by the observer gain $L$. As such, the power of the aggregate EV load, $\Delta PQ_{EV}$, is determined based on $\Delta PQ_{EV} = u = K_{def}\hat{x}$. Designing the EV-based controller based on feedback control law changes the A matrix to its closed-loop form, i.e., $A_{cl} = A + BK_{def}$. Since the eigenvalues of $A_{cl}$ define the stability of the power grid, it is of paramount importance that the methodology used to calculate $K_{def}$ guarantee the mitigation scheme's performance under different types of attacks when faced with the different uncertainties and obstacles of a real-life deployment. The mentioned sources of uncertainty in the EV load are discussed in the controller design section and studied in Section 5.

### 3.2. OBSERVER DESIGN

Since our EV-based LA attack mitigation strategy relies on full-state feedback control, the utility also needs to choose an appropriate design for the observer gain matrix $L$. This observer facilitates the accurate recovery and estimation of the grid states $\hat{x}$ needed to calculate the feedback control input $\Delta PQ_{EV} = u = K_{def}\hat{x}$. With the introduction of the gain $K_{def}$ and the observer $L$, the state-space (1) becomes (2).

$$\dot{\hat{x}} = A_{cl}\hat{x} + L(y - \hat{y}) + B_d\omega \quad (2a)$$
$$y = C_{cl}x + D_d\omega \quad (2b)$$

where $C_{cl} = C + DK_{def}$ is the closed-loop form of matrix C that relates the system outputs to its states. $B_{cl}$ and $D_{cl}$ are $B_d$ and $D_d$ respectively. Given that an accurate observer is based on gain matrix $L$, it must be designed to ensure that $(A_{cl} - LC_{cl})$ has stable poles [43]. Therefore, we employ a Linear Quadratic Controller (LQR) [43] as our observer for accurate state recovery. LQR is a control technique that optimizes the balance between the energy of states, and control signals to achieve the desired output accurately. This balance is controlled by the respective weights Q and R of the states and inputs. Q is a square symmetric positive semi-definite matrix and R is a square symmetric definite matrix. The observer gain $L$ is designed as an optimization problem having cost function J:

$$minimize\ J = \int_0^\infty x_o^T Q x_o + u_o^T R u_o\ dt \qquad (3)$$

where $x_o = (y - \hat{y})$ is the difference between the actual measured outputs and their estimated value, and $u_o$ is the observer output being fed back into the state-space. Given that $Q = Q^T \geq 0$ and $R = R^T > 0$, this problem can be solved by finding S that satisfies the Algebraic Riccati Equation (4)

$$0 = SA_o + A_o^T S - SB_o R^{-1} B_o^T S + Q \qquad (4)$$

where $A_o = A_{cl}^T$, $B_o = C_{cl}^T$ and $S = S^T \geq 0$. The detailed derivation of (4) from (3) is omitted for compactness. Equation (4) is quadratic in S and has no trivial solution, but it has a single positive definite solution that makes the observer stable. Thus, the observer gain $L$ is determined as

$$L = -R^{-1} B_o^T S. \qquad (5)$$

### 3.3. CHOICE OF CONTROLLER FOR EV-BASED MITIGATION SCHEME

Unlike the studies in [29]- [32] that support the grid's steady-state operation using EVs, we utilize EVs to create a mitigation scheme to attenuate the impact of LA attacks. Additionally, unlike traditional approaches that create controllers to mitigate specific disturbances, we use EVs to mitigate the three known types of LA attacks by optimizing the controller to accomplish multiple objectives. Also, by formulating our mitigation scheme as a feedback controller, we eliminate the need for attack detection tools since the controller reacts to the changes in the grid's states. These states are recovered by the utility by using the observer suggested above. We suggest a family of H-2, H-∞, and mixed H-2/∞ control techniques to synthesize our problem as a convex optimization with LMI constraints [44] to achieve a guaranteed attenuation level of the attack impacts.

H-∞ controllers minimize the maximum singular value of a function while H-2 controllers minimize the energy of the output signal over the entire frequency range. This would result in the H-2 controller performing better than H-∞ over most frequencies but failing at specific frequencies. H-∞ controllers have been adopted in mechanical systems such as missile or satellite trajectory control [45] and suspension systems [46]. However, we intend to adopt the usage of such controllers into the domain of power grid protection against LA attacks. Although, H-2 controllers have received less attention in the literature than H-∞, their ability to outperform H-∞ at most frequencies merits their usage in our study. Given the complexity of our problem and the need to address multiple objectives simultaneously, we ultimately choose the mixed H-2/∞ controller to enable our EV-based mitigation scheme to handle the different types of LA attacks.

### 3.4. H-2 AND H-∞ CONTROLLER DESIGN

In this subsection, the EV-based mitigation controller scheme is developed, and its mathematical formulation is obtained based on the LMIs of the desired control law [44]. Writing our controller equations as LMIs gives us the flexibility to (i) finetune their design, (ii) implement complex control schemes, and (iii) combine multiple objectives into a single optimization problem. In our study, the unknown LA attack is treated as a persisting disturbance, $w$, whose impact we aim to attenuate. The relationship between the disturbance and the system states and outputs is governed by (2) presented above. Our mitigation scheme aims to simultaneously minimize the impact of the attack and the EV load involved in the feedback controller signal. To this end, we design our control methods below and modify the design to achieve all our desired objectives in the sub-sections that follow.

1) H-2 controllers [44] aim to minimize the $L^2$ norm of a system across the entire frequency range. This means that the cost function of the H-2 controller is the Euclidean distance of the outputs from the origin. This allows the H-2 controller to rapidly react and eliminate the disturbances in a system by using rapidly increasing input signals. The transfer function $T_2$, representing the influence of the disturbance $\omega$ on the grid output, i.e., generator frequency $y$, is presented in (6):

$$T_2 = \frac{y}{\omega} = (C + DK_{def})(sI - (A + BK_{def}))^{-1}B_d \qquad (6)$$

The cost function of the H-2 controller becomes $||T_2||_{L^2} < \gamma$. This function is rearranged in terms of its LMIs (7)-(8).

$$(AX + BK_{def}X)^T + AX + BK_{def}X + B_d B_d^T < 0 \qquad (7)$$
$$\text{trace } \{(C + DK)X(CX + DK_{def}X)^T)\} < \text{trace}(Z) < \gamma^2 \qquad (8)$$

Based on Schur's formulation for partitioned matrices [44], inequality (8) is rewritten as (9) and (10).

$$\begin{bmatrix} -Z & CX + DK_{defX} \\ (CX + DK_{def}X)^T & -X \end{bmatrix} < 0 \qquad (9)$$

$$trace(Z) < \gamma^2 \qquad (10)$$

Considering that the gain $K_{def} = K_{H-2} = WX^{-1}$, (7) and (9) now become (13) and (14), and the optimization problem to reduce the disturbance impact can be written as (11)-(14).

$$\text{Minimize } \gamma^2 \qquad (11)$$
$$\text{s.t.} \qquad trace(Z) < \gamma^2 \qquad (12)$$
$$\begin{bmatrix} -Z & CX + DW \\ (CX + DW)^T & -X \end{bmatrix} < 0 \qquad (13)$$
$$(AX + BW)^T + AX + BW + B_d B_d^T < 0 \qquad (14)$$

This LMI optimization problem can only have a solution if there exists a matrix W and two symmetric matrices Z and X satisfying the matrix inequalities in (12)-(14).

2) H-∞ controllers [44] aim to minimize the maximum singular value of a function. This means that the H-∞ controller aims to minimize the largest perturbation in a system. The cost function of this controller is formulated to minimize the output and input energy meaning it eliminates the disturbances while simultaneously trying to minimize the required control inputs. This cost function can be expressed as the infinity norm, $||T_\infty||_\infty < \rho$, where $T_\infty$ is the transfer function dictating the influence of the disturbance $\omega$ on the grid output $y$.

$$T_\infty = \frac{y}{\omega} = (C + DK_{def})(sI - (A + BK_{def}))^{-1}B_d + D_d \qquad (15)$$

Based on Schur's formulation for partitioned matrices [44], the cost function $||T_\infty||_\infty < \rho$ is rewritten as an LMI and $K_{def} = K_{H-\infty} = WX^{-1}$ is replaced in the derived LMI. Then similar steps to (9)-(14) are taken to arrive at the inequality in (18) and the H-∞ optimization problem becomes (16)-(18).

$$\text{Minimize } \rho \tag{16}$$

$$\text{s.t.} \quad X > 0 \tag{17}$$

$$\begin{bmatrix} (AX+BW)^T + AX + BW & B_d & (CX+DW)^T \\ B_d^T & -\rho I & D_d^T \\ CX + DW & D_d & -\rho I \end{bmatrix} < 0 \tag{18}$$

The H-∞ LMI optimization problem can only have a solution if there exists a matrix W and a symmetric positive definite matrix X satisfying the matrix inequality problem in (17)-(18).

Each of these two control techniques is better suited to a specific class of LA attacks as we will demonstrate in Section 4 which is why we will also develop a mixed controller later in this section. After calculating $K_{def}$, the utility will have to check the attenuation levels dictated by $\gamma$ and $\rho$ respectively. These variables should be maintained below 1 to achieve feasible and satisfactory attenuation levels. This, however, only guarantees the performance of the controller in the frequency domain. To improve the controller performance in the time domain, i.e., reduce settling time, we add the pole placement constraint (19).

$$AX + XA^T + BW + W^T B^T + 2\Lambda_1 X < 0 \tag{19}$$

where $X$ is a positive semi-definite matrix. This LMI constraint is based on a D-stabilization pole placement technique [12] that shifts the eigenvalues of the system into a region where the real part of the eigenvalues is less than $-\Lambda_1$. To guarantee performance and fast settling time, we choose $l < \Lambda_1 < 0$ to ensure stable poles. $l$ should also be chosen small enough to avoid an aggressive controller behavior that is not desirable.

### 3.5. ROBUST CONTROLLER UNDER UNCERTAIN EV FEEDBACK LOAD

In this subsection, we address the issue of uncertainty of the EV load. This uncertainty can arise from different sources. The first is the user behavior that can be accurately estimated but never guaranteed by the utility. The second source of uncertainty arises from the possibility of attackers targeting the EV ecosystem. Attackers might be able to compromise part of the connected EVs while giving the utility the impression they are secure. This would mean that the control signal would reach less secure EVs than the utility intended. The third source of uncertainty can be introduced in the system by the clustering method we suggest below as a privacy-preserving measure. Clustering is the only case where uncertainty can be positive. These three types of uncertainties are modeled as an uncertain matrix $\emptyset$ in the feedback loop after the output of the controller $K_{def}$ which changes the value of the EV load to $\Delta PQ_{EV-\emptyset} = u_\emptyset = \emptyset K_{def}\hat{x} = \emptyset u$. As a result, the state-space representation in (1) becomes (20).

$$\dot{x} = Ax + B\emptyset K_{def}\hat{x} + B_d\omega \tag{20}$$

By rewriting $\Delta PQ_{EV-\emptyset}$ as a function of the original input $u$, (20) becomes (13) which is restructured through (22) to become the state-space representation in (23) which models the uncertainty in the feedback signal as uncertainty in matrix $B$.

$$\dot{x} = Ax + B\emptyset u + B_d\omega \tag{21}$$
$$\dot{x} = Ax + \{B + B(\emptyset - I)\}u + B_d\omega \tag{22}$$
$$\dot{x} = Ax + (B + \Delta B)u + B_d\omega \tag{23}$$

It is worth noting that since the utility is responsible for this mitigation scheme, it is considered that the system parameters are accurate, thus the uncertainty in matrix A is zero. The matrix uncertainties are written as (24) and (25).

$$\Delta A = HFE_1 \qquad (24)$$
$$\Delta B = HFE_2 \qquad (25)$$

Where H, $E_1$, and $E_2$ are known quantities while F is an uncertain matrix. Since $\Delta A$ is zero, then $E_1$ is zero and excluded from further calculations. H is usually chosen to be an identity matrix. The uncertain matrix F can be written in the form of (26) if it satisfies condition (27).

$$F = \delta_1 F_1 + \delta_2 F_2 + \cdots + \delta_k F_k \qquad (26)$$
$$FF^T \leq I \qquad (27)$$

Writing F in the form of (26) means that $\Delta B$ can be rewritten as (28) and consequently as (29) which represents a family of uncertain matrices.

$$\Delta B = H(\delta_1 F_1 + \delta_2 F_2 + \cdots + \delta_k F_k) E_2 \qquad (28)$$
$$\Delta B = \delta_1 B_1 + \delta_2 B_2 + \cdots + \delta_k B_k \qquad (29)$$

The H-2 controller constraint (14) is now rewritten as (30) and simplified as (31) to account for the uncertainty in B.

$$\{AX + (B + \Delta B)W)\}^T + AX + (B + \Delta B)W + B_d B_d^T < 0 \qquad (30)$$
$$(AX + BW)^T + AX + BW + B_d B_d^T + \beta < 0 \qquad (31)$$
$$\text{where } \beta = \Delta BW + (\Delta BW)^T \qquad (32)$$

Based on (25)-(29), $\beta$ is rewritten as (33) and (34).

$$\beta = HFE_2 W + (HFE_2 W)^T \qquad (33)$$
$$\beta = HFE_2 W + W^T E_2^T F^T H^T \qquad (34)$$

Using the variable elimination Lemma in [44] that states that any variable in the form of (34) under the condition stated in (27) can be rewritten as (35) $\beta$ becomes:

$$\beta = \alpha HH^T + \alpha^{-1} (E_2 W)^T (E_2 W) \qquad (35)$$

if there exists a scalar $\alpha > 0$. After substituting the value of $\beta$ derived in (35) back into (31), we apply Schur's complement lemma [47], to rearrange the inequality containing $\alpha^{-1}$ into the equivalent inequality in (36) which would replace constraint (14) in the original H-2 formulation making it a robust H-2 controller capable of handling uncertainty in the feedback loop.

$$\begin{bmatrix} \Lambda_1 + \alpha HH^T & (E_2 W)^T \\ E_2 W & -\alpha I \end{bmatrix} < 0 \qquad (36)$$
$$\text{where } \Lambda_1 = (AX + BW)^T + AX + BW + B_d B_d^T. \qquad (37)$$

Without going into the details, similar steps to (25)-(35) are followed and the inequality in (18) is rewritten as (38)

$$\begin{bmatrix} \Lambda_2 + \alpha HH^T & B_d & (CX + DW)^T & (E_2 W)^T \\ B_d^T & -\rho I & D_d^T & 0 \\ CX + DW & D_d & -\rho I & 0 \\ E_2 W & 0 & 0 & -\alpha I \end{bmatrix} < 0 \qquad (38)$$
$$\text{where } \Lambda_2 = (AX + BW)^T + AX + BW \qquad (39)$$

This formulation represents a robust H-∞ controller capable of handling uncertain parameters represented by $\Delta B$ corresponding to our feedback loop EV load uncertainty.

### 3.6. MIXED ROBUST H-2/∞ CONTROLLER

After presenting the robust controller formulation above, we discuss herein the robust mixed H-2/∞ controller. Our mitigation strategy ultimately aims at eliminating the impact of the three types of LA attacks hence the importance of its success in the different attack ranges and the maximum possible attenuation level across these ranges. To this end, we develop a mixed controller that combines the LMI constraints of both H-2 and H-∞ robust controllers and arrive at the formulation in (40)-(47). The optimization objective function (40) is a weighted mix of objectives (11) and (16).

$$\text{Minimize } a_1\gamma^2 + a_2\rho \tag{40}$$

$$\text{s.t.} \quad AX + XA^T + BW + W^TB^T + 2\Lambda_1 X < 0 \tag{41}$$

$$\begin{cases} X > MI & (42) \\ \text{trace}(Z) < \gamma^2 & (43) \\ \begin{bmatrix} -Z & CX + DW \\ (CX + DW)^T & -X \end{bmatrix} < 0 & (44) \\ \begin{bmatrix} \xi + B_d B_d^T + \alpha HH^T & (E_2W)^T \\ E_2W & -\alpha I \end{bmatrix} < 0 & (45) \\ \begin{bmatrix} \xi + \alpha HH^T & B_d & (CX + DW)^T & (E_2W)^T \\ B_d^T & -\rho I & D_d^T & 0 \\ CX + DW & D_d & -\rho I & 0 \\ E_2W & 0 & 0 & -\alpha I \end{bmatrix} < 0 & (46) \end{cases}$$

$$\text{where } \xi = (AX + BW)^T + AX + BW \tag{47}$$

Constraint (17) is replaced by $X > MI$ when incorporated into the mixed controller where M is a positive integer. This new constraint guarantees a smaller value of the gain $K_{def}$ consequently, a smaller input signal, $u = \Delta P Q_{EV}$, to be sent to the EV-based defender scheme. This mixed sensitivity LMI optimization problem can only have a solution if there exists a matrix W, two symmetric matrices Z and X, and a scalar $\alpha$ satisfying (40)-(47) and a gain matrix $K_{def} = K_{\text{mix}} = WX^{-1}$.

### 3.7. EV USER IMPACT AND PRIVACY CONCERNS

The main concern related to the suggested scheme is the impact on the EV users' charging experience since such a scheme would require some of the connected EVs to charge and/or discharge. However, the average charging time for 100 miles of range is 20 hours per 1 kW charging rate [48]. This means that 11 kW EVCSs would deliver a charge equal to 0.9 miles/min, which varies between different EVs. The impact on EV range is thus minimal and discussed in Section 4.

Such a scheme should only involve EVs whose owners gave their consent to participate. This can be achieved through an incentive program implemented by the utility to motivate EV users to contribute to the mitigation scheme. Hydro-Quebec, a Canadian Utility, implements a similar program, the *Hilo* project [49], which allows them to control home loads to reduce the demand during peak times with plans to extend it to EVs.

Regarding private EVCSs at residences, the issue of user privacy comes to light. The trivial solution would be to implement this scheme on public EVCSs only. However, given the abundance of home EVCSs, we suggest an EVCS clustering by the utility based on geographical location. Utilities can broadcast a signal to the entire cluster without having granular knowledge of individual home EVCSs. For example, let's consider areas A and B with 500 and 350 EVCSs respectively. Also, let's assume that the estimated EV connection to the EVCSs in area A and area B is 20% and 30% respectively. If the utility wants to cluster the EVCSs such that 10 EVs are connected per cluster, this requires area A to have 10 clusters of 50 EVCSs, and area B to have 10 clusters of 35 EVCs. This, however, introduces uncertainty in the actual EV load used as a feedback control signal in the mitigation scheme since the number of connected EVs per cluster cannot be guaranteed. The impact of this clustering method is examined in Section 5 as part of our EV-based mitigation scheme's stability study.

## 4. CASE STUDIES AND SIMULATIONS

In this section, we demonstrate our EV mitigation scheme against static, switching, and dynamic attacks on the New England (NE) 39-bus grid in Fig. 2. [50]. We first demonstrate the impact of the attacks in the absence of our scheme to highlight their devastating impact. The NE grid has 39 buses, 10 generators, and 19 loads with a total of 6,097MW. The simulations were performed on MATLAB-Simulink 2021a Specialized Power Systems Toolbox using a variable step size of $1 \times 10^{-12}$ to $1 \times 10^{-9}$.

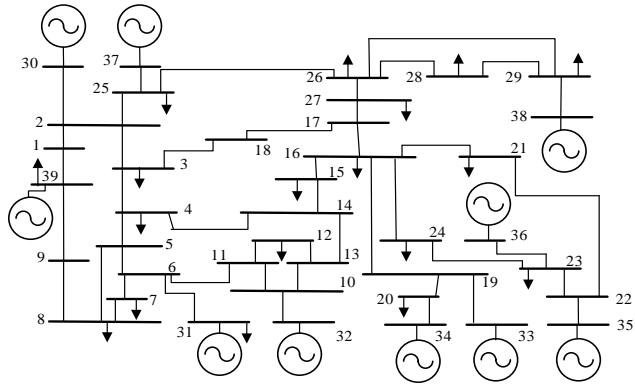

Fig.2. New England 39-bus grid

While we acknowledge that the current number of EVs is not enough to exploit the full potential of the suggested mitigation scheme, the following example demonstrates its feasibility with increased EV adoption levels. To demonstrate this, we choose a similarly sized grid which is the New South Wales (NSW) grid on a day in December 2021 [51]. The average load is 6968 MW [51] and the total registered vehicles are 5,892,206 [52]. Scaled down to fit our test grid, the total number of vehicles is 5,155,681.

At future EV penetration levels, the EV load will become huge, especially with the increasing EVCS charging rates. Thus, only a small fraction of the EVs in our grid needs to be connected for our proposed EV-based LA attack mitigation scheme to achieve the results presented below. However, the EV load is variable depending on the time of day. Thus, a more detailed examination of the EV load is required. As per the International Energy Agency (IEA) [53] EVCS operators, utilities, and governments have maintained a ratio of 1 public EVCS for every 9.9 EVs on the road to guarantee the quality of service. Thus, at a future 50% EV penetration level, there will be 260,387 public EVCSs in our NE grid. With an average public EVCS occupancy of 33% [54], we can estimate the average number of connected EVs to the grid to be 86,000. Furthermore, according to the IEA [53], based on the mixture of different EVCS rates, the average charging rate per EVCS is 24kW. From these statistics, we can estimate that at a future 50% EV penetration level, there will be 2,064MW of EV load connected to the grid on average at public EVCS.

To examine the specific EV load during the different times of the day, we create a data-driven model for the arrival and charging times of the EVs. To achieve a realistic EV charging behavior, we independently simulate a Poisson arrival process of EVs to each EVCS [55]. The charging time of these EVs is assumed to follow a truncated Gaussian distribution [55]. The parameters of these models are specified for 1-h windows for a 24-h period. These parameters are tuned based on a real dataset containing five years of records for 7,500 EVCSs in Quebec, Canada. This dataset was obtained from Hydro-Quebec as part of a research collaboration.

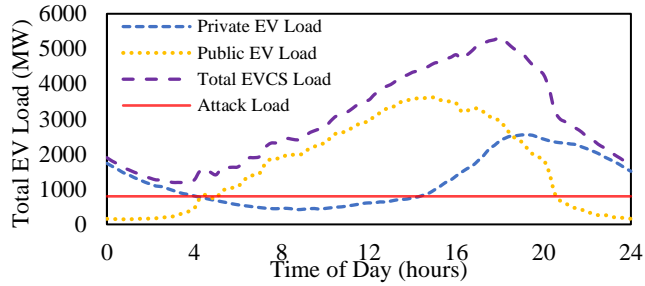
Fig. 3. EVCS Public and Private charging load for 24 hours

Additionally, we extract the EVCS utilization information from the dataset. The average utilization rate of EVCSs in Quebec is 31% with the peak charging demand occurring in the afternoon. By examining this dataset, we were able to extract the average hourly arrival rates and charging times to simulate them as a Poisson process and a truncated Gaussian distribution respectively. From the presented statistics and data-driven EV fleet model, we generate the public EVCS load profile presented in Fig. 3. Additionally, we acquire an approximation of the private EVCS load profile, for the presented number of EVs, by using the Electric Vehicle Infrastructure Projection Tool provided by the Alternative Fuels Data Center (AFDC) [56] and added the acquired data to Fig. 3. Fig. 3 demonstrates the change in the EV load for an entire 24-h period. Fig. 3 demonstrates the presence of a minimum 1196MW of EV charging load connected to the grid at 3:30 am.

Furthermore, there are roughly 9 private EVCSs for every 10 EVs [53]. This means that at 50% penetration, there will be over 2.3 million private EVCSs. Given user tendency in the presence of an incentive scheme like *Hilo*, to connect EVs to home EVCSs even if no significant charging is needed, there will be large numbers of EVs connected at residences at the disposal of the utility that do not factor into the normal charging load presented in Fig. 3. Once we factor in the EVs that will be connected due to the incentives, the utility will have a much larger EV charging load at its disposal to participate in the presented mitigation scheme.

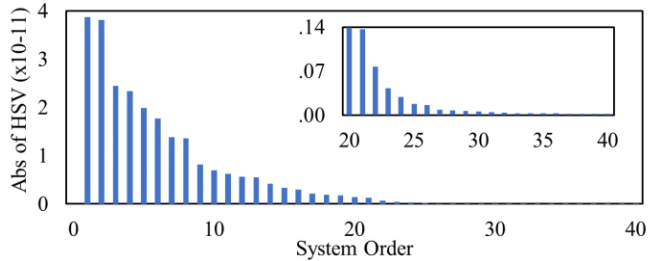
Fig. 4. First 40 Hankel Singular Value Weights

### 4.1. REDUCED POWER GRID STATE-SPACE MODEL

Due to the complexity of power grids, the size of the state-space matrices is relatively large hindering the design of the controllers, i.e., the NE grid has 300 states. To preserve the correct system behavior while reducing its order, we use Hankel model reduction [57]. For this purpose, we calculate the Hankel Singular Values (HSVs) of the system states to optimize the system order reduction versus the preserved system accuracy. Fig. 4 represents the energy contained within the Hankel Singular Values (HSVs) of the first 40 states. Fig. 4 shows the sharp decrease in the energy of the HSVs as the order of the system states increases. Table 1 shows the preserved system energy based on the reduced system order. Since 99.81% of the total system energy is contained within the first 30 HSVs, we reduced the system order to 30 states. Beyond 30, any order increase does not significantly improve accuracy while considerably adding to the controller synthesis complexity. This order reduction is only used for controller synthesis, while the presented simulations are performed on the actual grid, not the linearized nor the reduced model.

TABLE 1. POSSIBILITIES FOR REDUCING SYSTEM ORDER

| Reduced System Order | Percentage of Preserved System Energy |
|---|---|
| 10 | 83.93% |
| 20 | 98.38% |
| 30 | 99.81% |
| 40 | 99.93% |
| 50 | 99.97% |

**4.2. CASE STUDY 1: ATTACKS IN ABSENCE OF MITIGATION WITHOUT PSS**

Based on the work presented in [7], [10], and [12] which discuss static attacks, switching attacks, and dynamic attacks respectively, we formulate our attack scenarios against the NE grid in the absence of the suggested mitigation strategy to demonstrate their impact as a base case. The mathematical formulations of these attacks are presented in [7], [10], and [12], and are based on these studies, thus are not repeated herein. The first batch of attacks is simulated in the absence of a PSS and the impact is discussed in the following three cases.

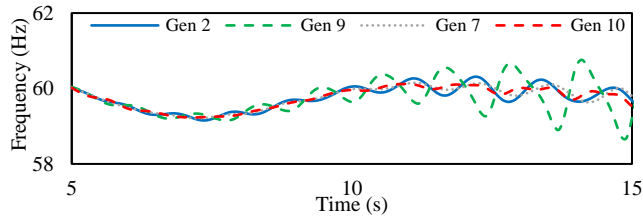
Fig. 5. Generator frequency response under attack 1 with no PSS

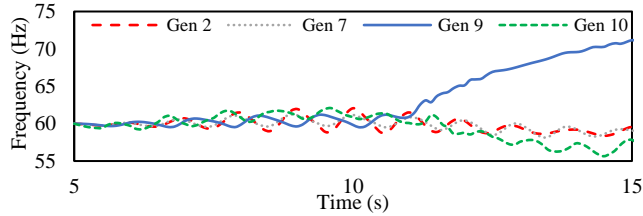
Fig. 6. Generator frequency response under attack 2 with no PSS

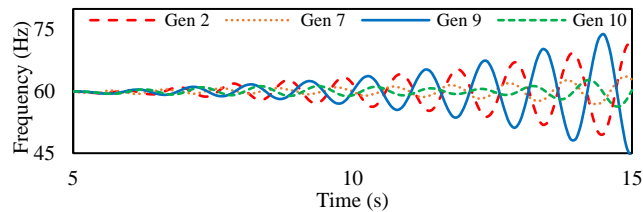
Fig. 7. Generator frequency response under attack 3 with no PSS

1) Attack 1: is an 800 MW static attack initiated at t=5s split on buses 3, 4, 24, and 29. This attack is a single spike in load equal to 13% of the system demand. Fig. 5 demonstrates the frequency behavior of the generators in response to this attack. Attack 1 causes a frequency drop to 59 Hz followed by growing oscillations that are sustained due to the absence of a PSS.

2) Attack 2: is an 800 MW switching attack split equally on buses 3, 4, 24, and 29. Fig. 6 demonstrates how this attack caused a huge frequency deviation that reached 71 Hz in 10 s. This attack would trip the generators resulting in a blackout.

3) Attack 3: is an 800 MW dynamic attack split on buses 3, 4, 24, and 29. Since the attack is formulated as a feedback loop so that the eigenvalues of the grid are shifted into an unstable region based on the attack methodology in [12], the loads do not always oscillate in phase. The result of this attack is demonstrated in Fig. 7. The frequencies of all the generators experience wild oscillations that continue to grow to reach 74 Hz while oscillating against each other. This would trip all the generators to avoid damaging their shafts due to the violation of the safe frequency operation thresholds.

**4.3. CASE STUDY 2: ATTACKS IN ABSENCE OF MITIGATION WITH PSS**

In this case study, we repeat the three attack scenarios in the presence of the PSS at the generators. The PSS is meant to stabilize the grid by damping generator frequency swings but falls short of eliminating LA attack impacts.

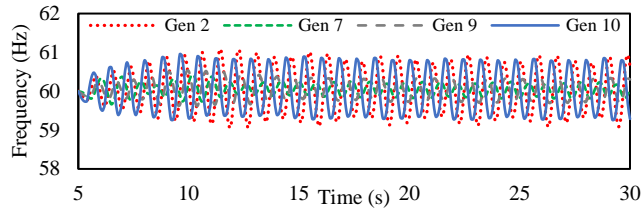
Fig. 8. Generator frequency response under attack 2 with PSS

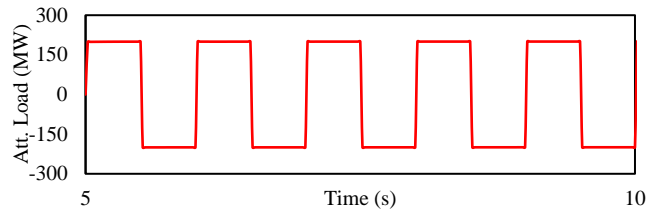
Fig. 9. Attack 1 load on each of the buses

1) Attack 1: is now repeated in the presence of the PSS. This static attack causes a 0.4 Hz frequency drop and minor oscillations that are eventually eliminated by the PSS as the grid regains stability.

2) Attack 2: is also repeated in the presence of the PSS which significantly limits the impact. However, the grid sustained frequency oscillations reaching a maximum deviation of 1.08 Hz. Fig. 8 demonstrates this impact that causes the operator to shed 5% of the grid's load [8]. Fig. 9 demonstrates a sample Attack 2 load. The average frequency deviation among the 10 generators was 0.74 Hz.

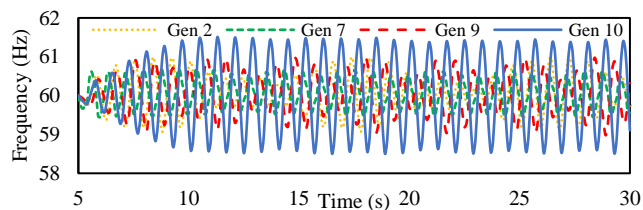
Fig. 10. Generator frequency response under attack 3 with PSS

3) Attack 3: is also repeated in the presence of the PSS and again proved to be the strongest attack. The grid experienced the rapid frequency oscillation depicted in Fig. 10. The maximum frequency reached 1.58 Hz. This causes the generators to trip instantaneously leading to a blackout. Fig. 11 demonstrates a sample Attack 3 load and Fig. 12 represents the aggregate of the 4 attack loads. The average frequency deviation among the 10 generators was 1.06 Hz.

This case study demonstrates that even in the presence of the PSS the attack impact is not eliminated when it is crafted properly. To this end, adding a mitigation mechanism that can eliminate LA attack impacts becomes necessary.

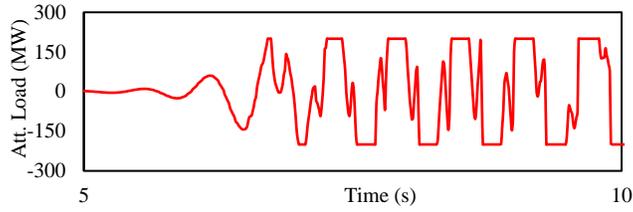
Fig. 11. Attack 3 load on bus 4

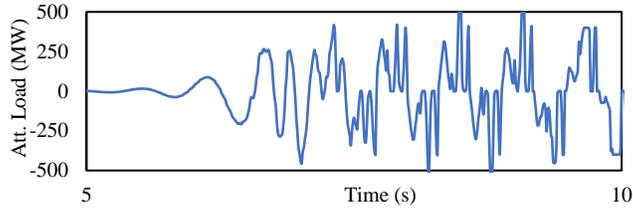
Fig. 12. Attack 3 aggregate load

### 4.4. CASE STUDY 3: H-2 AND H-∞ CONTROL EV-BASED MITIGATION

Based on the methodology presented above, we design our EV-based mitigation scheme based on H-2 and H-∞ robust control and evaluate the performance of both controller designs against the three types of LA attacks. The EV defender load is calculated as $\Delta PQ_{EV} = K_{def}\hat{x}$ and replicated on the NE grid that is under attack. It is important to mention that the H-2 and H-∞ EV-based mitigation, immediately eliminates all traces of the static attack. The cases of switching and dynamic attacks are presented below.

1) H-2 EV mitigation: We repeat Attack 2 and Attack 3 on the NE grid after adding the EV feedback loop gain, $K_{def}$, calculated based on H-2 control and the results are as follows.

- Attack 2: is repeated and the frequency responses of the generators are presented in Fig. 13. The maximum frequency deviation was reduced to 0.065 Hz which represents a 93.9% decrease in the attack impact. The average frequency deviation was reduced to 0.04 Hz which is equal to a 99.6% reduction in attack impact.

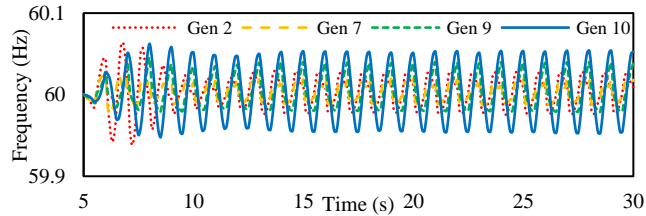
Fig. 13. Generator frequency - attack 2 with H-2 controller

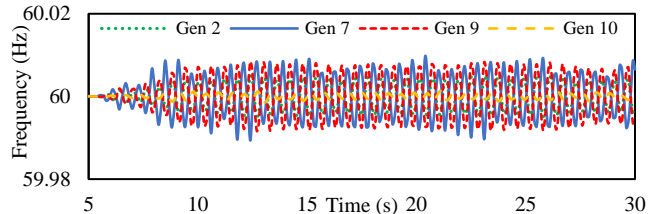
Fig. 14. Generator frequency - attack 3 with H-2 controller

- Attack 3: is repeated and the frequency response of the generators is tremendously improved. As presented in Fig. 14, the maximum frequency deviation was reduced to 0.007 Hz which is roughly equivalent to a 99.6% reduction in the original attack impact, practically eliminating it. The average frequency deviation was reduced to 0.004 Hz which is equal to a 99.6% reduction in attack impact.

2) H-∞ EV mitigation: We repeat Attack 2 and Attack 3 on the NE grid after adding the EV feedback loop gain $K_{def}$ calculated based on H-∞ control and the results are as follows.

- Attack 2: is repeated and causes a frequency drop of 0.01 Hz which then recovers at t=30s to 60 Hz, representing 100% recovery to the pre-attack state as seen by the frequency responses of the generators in Fig. 15.
- Attack 3: is repeated in the presence of the H-∞ EV mitigation scheme. Although this controller eliminates the impacts of the switching attacks, it falls short of achieving the same against the dynamic attack. The frequency responses of the generators are presented in Fig. 16. The maximum frequency deviation was reduced to 0.1 Hz which is a 93.7% reduction of the LA attack impact. The average frequency deviation was reduced to 0.05 Hz which is equal to a 95.3% reduction in attack impact.

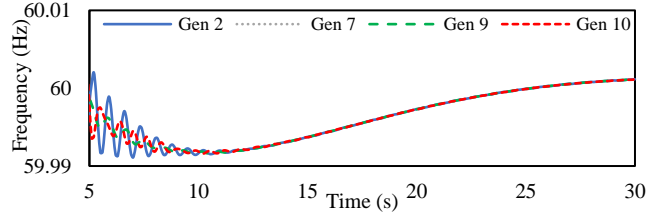
Fig. 15. Generator frequency - attack 2 with H-∞ controller

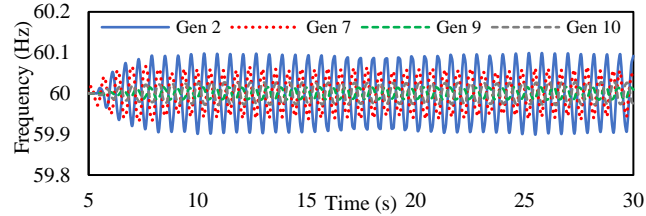
Fig. 16. Generator frequency - attack 3 with H-∞ controller

This case study demonstrates that the H-∞ EV-based mitigation scheme performs better against switching attacks while the H-2 mitigation scheme performs better against dynamic attacks. In the case of the switching attack in the presence of the H-2 and the dynamic attack in the presence of the H-∞ controller, the governor of the generators would have to react since the frequency does return to the safe operation limit but not to the normal frequency range. Since this is a persisting attack, this means that the governors will have to constantly keep correcting the frequency. As a result, we recommend a mixed controller.

We now repeat the attacks to test the mitigation scheme's success when the defender does not control any EVs on one of the attacked buses. This is to demonstrate the success of our EV mitigation even if we eliminate one of its advantages which is the colocation with the attack load. The controller is successful in eliminating the attack impact but with slightly reduced performance. Attack 2 is repeated against the grid that has the H-∞ controller and results in a maximum 0.1Hz frequency deviation which is equivalent to a 90.7% reduction in impact. The average frequency deviation was reduced to 0.04 HZ which is equivalent to a 94.6% reduction. Attack 3 is also repeated against the grid that has the H-2 mitigation scheme which results in a maximum 0.06 Hz frequency deviation or 96.2% reduction in attack impact. The average frequency deviation was reduced to 0.03 Hz or a 97.2% reduction. This demonstrates the success of our attack mitigation even when the utility loses its resources on an attacked bus.

### 4.5. CASE STUDY 4: ROBUST MIXED H-2/∞ EV-BASED MITIGATION

In this case study, we demonstrate the effectiveness of the mixed H-2/∞ robust control mitigation strategy and its advantage in our EV-based LA attack mitigation. Once again, the H-2/∞ controller eliminates any trace of the static attack.

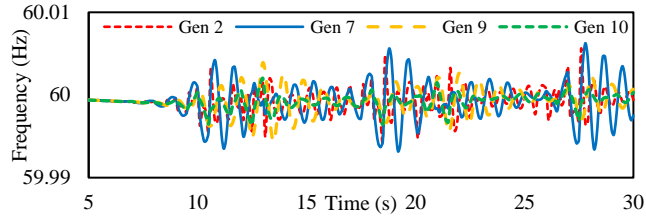
Fig. 17. Generator frequency - attack 2 with mixed H-2/∞ controller

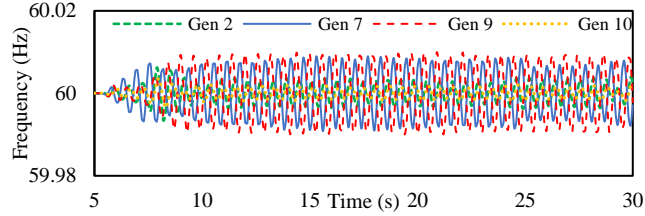
Fig. 18. Generator frequency - attack 3 with mixed H-2/∞ controller

1) Attack 2: is repeated and Fig. 17 demonstrates the success of the mixed control strategy in mitigating the impact of the switching attack. The frequency responses of the generators are presented in Fig. 17, and it is evident that the performance is much better than the H-2 controller with a reduction of the switching attack impact by 99.5%. The average frequency drops slightly below 60 Hz but stabilizes towards t=30s. Also, the sustained oscillations reach a maximum of 0.006 Hz. The average value of the sustained oscillations across all generators was also reduced to 0.003Hz representing a 99.6% drop in average attack impact.

2) Attack 3: is repeated and the frequency responses of the generators are presented in Fig. 18. Fig. 18 demonstrates the success of the controller in counteracting the impact of the dynamic attack. The maximum impact of the attack was reduced from 1.5 Hz to 0.01 Hz representing a 99.4% impact reduction. This is also 10 times smaller than the impact of the same attack in the presence of the H-∞ EV-based controller. The average value of the sustained oscillations was also reduced to 0.008 Hz representing a 99.3% drop in average attack impact.

3) We also study the case when the utility has no resources on the attacked buses. The switching attack and dynamic attack cause a maximum frequency deviation of 0.09 and 0.07 respectively. This is equivalent to a 91.7% and a 95.6% reduction in maximum attack impact respectively.

This case study proves that the mixed H-2/∞ controller is superior to the individual controllers by addressing their gaps. By reducing the frequency oscillation and deviation caused by all types of LA attacks to lower than 0.01 Hz, the mixed controller returned the grid to a state where the frequency is well within the normal range in which the turbine governors are not engaged, and the system behaves as it would behave normally in the absence of any attack even when the attacks are sustained and persistent. Based on the above discussions and results, the best course of action would be the adoption of the EV-based robust mixed H-2/∞ controller for the LA attack mitigation scheme. The complexity of designing the controller based on the presence of uncertainty is only during the planning phase. During deployment, the performance of the controller will not be impacted since it is based on matrix multiplication regardless of the method of its synthesis. Lastly, the presented case studies demonstrate the instantaneous reaction time of the presented mitigation scheme due to the advantage introduced by the power converters in the EVCSs that resulted in the immediate elimination of the attack impact.

### 4.6. CASE STUDY 5: SMALLER LA ATTACKS

The previous case studies aimed at demonstrating the EV-based robust mitigation scheme's success against large LA attacks (800MW) to showcase the mitigation scheme's effectiveness. In this case study, however, we examine the impact of smaller attacks and the performance of our proposed EV-based mitigation scheme in such scenarios. As a starting point, we repeat Attack 2 and Attack 3 and cap their attack loads at different magnitudes between 100MW and 800MW. Table 2 demonstrates the impacts of such attacks before and after the addition of our EV-based robust mixed H-2/∞ mitigation scheme.

TABLE 2. MAX FREQUENCY DEVIATION VS DIFFERENT ATTACK 2 AND 3 MAGNITUDES

| Attack Magnitude | Attack 2 | | Attack 3 | |
|---|---|---|---|---|
| | No Mitigation | Mixed H-2/∞ Mitigation | No Mitigation | Mixed H-2/∞ Mitigation |
| 800 | 1.08 Hz | 0.006 Hz | 1.58 Hz | 0.01 Hz |
| 700 | 0.97 Hz | 0.006 Hz | 1.42 Hz | 0.01 Hz |
| 600 | 0.87 Hz | 0.005 Hz | 1.21 Hz | 0.009 Hz |
| 500 | 0.79 Hz | 0.004 Hz | 1.03 Hz | 0.008 Hz |
| 400 | 0.65 Hz | 0.003 Hz | 0.91 Hz | 0.006 Hz |
| 300 | 0.54 Hz | 0.003 Hz | 0.76 Hz | 0.004 Hz |
| 200 | 0.32 Hz | 0.001 Hz | 0.46 Hz | 0.002 Hz |
| 100 | 0.17 Hz | 0.001 Hz | 0.26 Hz | 0.001 Hz |

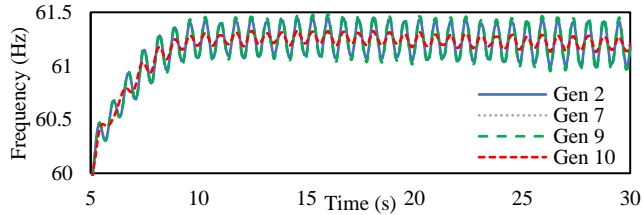

Fig. 19. Generator frequency - attack 4 with mixed H-2/∞ controller

We now examine a new dynamic attack (Attack 4) based on the methodology in [12] while shifting the eigenvalues of the system further right (unstable region) than Attack 3. This results in a faster oscillation of the attack load. Attack 4 is initiated at t=5s against buses 3, 4, 18, and 39 with a magnitude of 19% of the load on each bus for a total of 395.96MW. Attack 4 causes the average frequency to reach 61.25 Hz while the forced oscillations reach 61.49 Hz as depicted in Fig. 19. While this behavior does not cause instantaneous generator tripping since it does not exceed the 1.5 Hz limit, sustaining it for 30s will cause the generator protection relays to trip. However, some utilities have stricter limits (61Hz) meaning such an attack would instantaneously trip the generators. Attack 4 is now repeated in the presence of the mixed H-2/∞ mitigation strategy. The frequency deviation/oscillation is reduced below 0.01 Hz meaning that the attack impact was successfully eliminated.

### 4.7. IMPACT ON EV RANGE

We now evaluate the impact of our mitigation scheme on the EVs' range. This evaluation is based on the average charging rate of 24kW. Mitigating Attacks 2 or 3 requires the EV to alternate between charging and V2G such that the net charge is approximately 0. Attacks 2 and 3 cause a loss of 0.001kWh and 0.009kWh respectively. The EVs also lose the opportunity to charge 0.2kWh. The total loss is equivalent to a range of 1 mile for each EV. For an EV that was connected to an EVCS but not charging, the net impact is almost 0kWh. Attack 1 on the other hand will result in a total loss of 0.4kWh or 2 miles.

### 4.8. MITIGATION SCHEME FEASIBILITY

An added advantage of using EV charging in our mitigation scheme is that the required communication and control infrastructure is already in place. The required central authority needed to communicate with the distributed resources (EVCSs) is the CMS that already exists in the EV ecosystem. This CMS communicates and controls all the public EVCSs in real-time. Using the OCPP protocol, the CMS can turn the individual EVCSs on and off, change their charging rate, and discharge using V2G [25] [26]. This means that adopting our mitigation scheme would not require the addition of any software, hardware, or communication to the ecosystem. The Hilo project gives the utility the same control capabilities over private EVCSs. This means that adopting our mitigation scheme would not require any software and communication capabilities to be added to the ecosystem. Furthermore, since our mitigation scheme only requires the frequencies at the generators which are already monitored by the utility, implementing our mitigation scheme does not require the addition of any measurement devices.

Mitigating these attacks using our suggested scheme requires an extremely minimal cost to be incurred by the utility. To demonstrate this, we consider 2 different EVCS charging levels in Quebec with a charging rate of 24 and 50kW [58]. The hourly price of charging on these EVCSs is 7.53 CAD and 12.77 CAD respectively and is billed per second [58]. This means that mitigating Attack 2 and Attack 3 costs the individual EV user 6.28 cents on the 24kW EVCS and 5.1 cents on the 50kW EVCS. This means that when the utility reimburses the EV users for this cost, the utility will have to pay a total of 1,683-2,072 CAD (1,242- 1,530 USD) plus whatever extra value the utility determines in its incentive program. Mitigating Attack 1 however, requires double the cost since the energy loss from the EV batteries will be double that of the other 2 attacks.

### 4.9. MITIGATION SCHEME REACTION TO NON-ATTACK SCENARIOS

One issue that arises from the absence of a detection scheme is the EV controller's reaction to frequency fluctuations that are not caused by attacks. Our proposed scheme can react to sudden changes in power grid behavior such as the abrupt line or generator tripping studied in [15]. Such sudden events resemble static attacks. Fig. 20 demonstrates the grid's behavior after the line connecting bus 4 to bus 5 tripped in the presence of the H-2/∞ controller. It is evident that our EV- based mitigation successfully brings the grid back to stability after such a singular event and eliminated all traces of the impact.

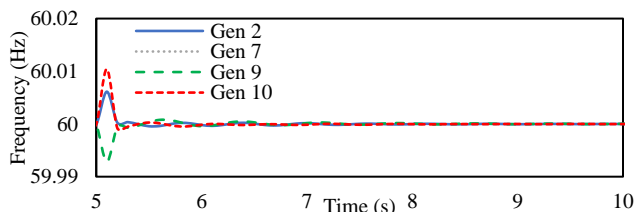
Fig. 20. Grid frequency after line tripping line 4-5

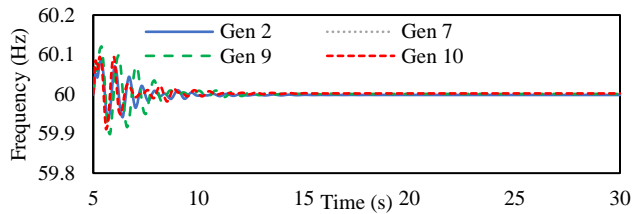
Fig. 21. Grid frequency after Generator 1 tripping

Additionally, Fig. 21 demonstrates the frequency behavior in the grid after tripping the generator connected at bus 39. This generator has an output of 1104MW, which is the largest in our grid. In the absence of our proposed mitigation scheme, the frequency deviation surpasses 1Hz and the generators start oscillating wildly leading to tripping. However, our proposed EV-based H-2/∞ controller immediately limits the initial deviation to 0.1 Hz and then brings the frequency back to the nominal 60 Hz.

These 2 simulation results demonstrate the effectiveness of the proposed EV-based LA attack mitigation scheme against the singular events that are usually studied in the literature in the context of EV frequency support to the grid. However, such events are rare during the normal operation of a power grid. To avoid having the presented EV-based mixed H-2/∞ controller react to the normal frequency fluctuations corresponding to the random consumer behavior, we set a frequency deviation threshold of 0.03 Hz before the mitigation scheme is engaged. This threshold can be changed by the utility depending on their historic data which would indicate the maximum value of benign frequency fluctuations caused by the random user behavior.

To simulate normal frequency fluctuations of a real grid, we add random load blocks to all load buses. IEEE benchmark grids have constant loads representing the average load of the individual buses. Utilities depend on historical data to estimate the load at a certain time of day. However, during short windows (ex.1min), the load cannot be predicted since real consumer behavior is random but centered around the average bus load. This gives rise to the need to simulate random perturbations in the loads of our grid which would lead to the normal frequency variations seen in Fig. 22. The random load variations follow a random Gaussian distribution in our simulations. To avoid the repetitiveness of pseudorandom number generator patterns (pattern effect) [59], we use the Mersenne Twister algorithm with a period of $2^{19937} - 1$ which can overcome the pattern effect [59] and guarantee true randomness. Additionally, we shuffle the random generator's seed before each simulation. After setting up this simulation environment, we simulated 3 weeks (21 days) of power grid behavior and collected the frequency readings. We then collected the value of the highest frequency deviation during each 1-minute window. A histogram representing the frequency deviation probability during normal behavior is presented in Fig. 23. This histogram demonstrates that normal frequency fluctuation caused by consumer behavior does not exceed 0.022 Hz. Thus, our threshold for engaging the mitigation scheme was set at 0.03 Hz.

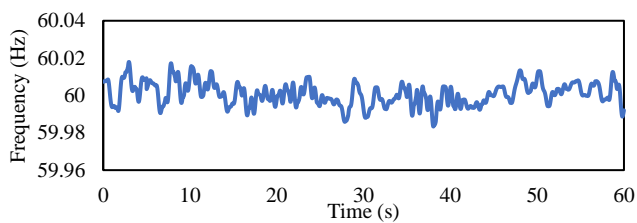

Fig. 22. Grid frequency under normal random consumer behavior

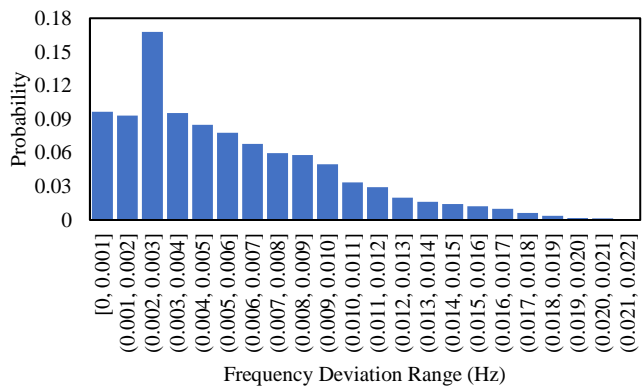

Fig. 23. Frequency deviation probability during normal behavior

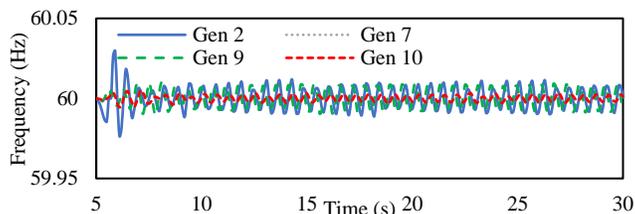

Fig. 24. Grid frequency under attack 3 in the presence of a 0.03 Hz threshold

In the following, simulation, we repeat Attack 3, which was the most impactful attack, in the presence of the mixed H-2/∞ controller after setting its engagement threshold to 0.03 Hz. Fig. 24 demonstrates the success of our mitigation strategy. The frequency rises to reach the 0.03 Hz threshold before the mitigation scheme is engaged and brings it back down to 0.01 Hz.

Additionally, our proposed scheme is superior to the EV-based frequency regulation schemes proposed in [33] - [35] since it is originally designed to deal with continuous persisting attacks especially the dynamic attack that adapts to the grid reaction. Such singular events resemble Attack 1, whose impact was eliminated completely. The work in [35] proposed a frequency regulation scheme for the small frequency fluctuations (<0.06 Hz) caused by the intermittency of renewable energy. Additionally, the presented scheme in [35] was optimized to handle sudden drops or spikes in renewable energy generation of 120MW resembling the behavior of Attack 1. The presented methods in [33] [34] on the other hand are meant to address frequency fluctuations between 0.06 Hz and 0.07 Hz, caused by renewable energy intermittency, and reduce the frequency deviation to 0.05 Hz (29% reduction). In comparison, our EV-based LA attack mitigation scheme is designed to handle persistent static, switching, and dynamic attacks. Additionally, our proposed controller is designed to mitigate the impact of attacks with much larger magnitudes than the events in [33] - [35] and reduce their impact by over 99% from the devastating range of 1.5 Hz to a normal 0.01 Hz range.

## 5. STABILITY AND PERFORMANCE EVALUATION

In this section, we examine the performance of our EV-based mitigation when faced with real-life deployment obstacles.

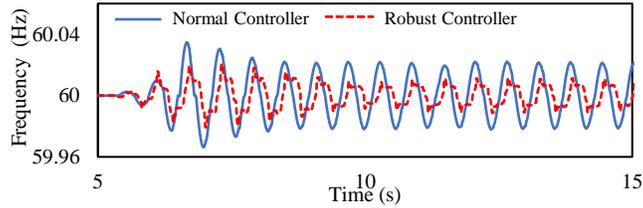

Fig. 25. Generator frequency with normal and robust controllers

### 5.1. UNCERTAINTY AND CONTROLLER STABILITY

In this subsection, we examine the stability of the proposed control strategy in face of the uncertainty of the feedback EV defender load. By modeling this uncertainty into the design of the controller, the value of the matrix $K_{def}$ changes to allow the mitigation scheme to perform well facing these uncertainties. In case there is no uncertainty present, a carefully designed robust controller will achieve an extremely similar response to the original non-robust controller but will outperform it when uncertainty is added to the systems. Fig. 25 presents the response of the worst-performing generator to Attack 3 in the presence of a random uncertainty of ±5% on each of the feedback EV inputs. Fig. 25 demonstrates how the mixed robust controller performs better than a normal mixed controller when the feedback channel is not 100% certain.

The three types of robust controllers were evaluated using MATLAB's "diskmargin" function to calculate their disk stability [57]. To this end, we consider the system states, x, as the object of our analysis for the sake of having a system with 30 inputs and 30 states. Table 3 demonstrates the stability range of the system when all the inputs are varied independently with an uncertainty $F_j$ presented (48). In (48) $j$ is the index of the input/state and δ is a random complex number such that $|\delta| < 1$ represents the uncertainty in the channel gain and phase. Variable $a$ is the multiplicative gain of the uncertainty and σ represents the skewness of $F$ to shift the probability of the uncertainty in the positive or negative direction.

$$F_j = \frac{1 + a\{(1-\sigma)/2\}\delta_j}{1 - a\{(1+\sigma)/2\}\delta_j} \tag{48}$$

TABLE 3. CONTROLLER STABILITY MARGIN

| Skew | σ=-2 | σ=-1 | σ=0 | σ=1 | σ=2 |
|---|---|---|---|---|---|
| Controller | min/max | min/max | min/max | min/max | min/max |
| Robust H-2 | 0.01/1.50 | 0.01/2.00 | 0.18/5.56 | 0.53/8.34 | 0.67/9.08 |
| Robust H-∞ | 0.20/1.44 | 0.29/1.70 | 0.41/2.17 | 0.61/2.62 | 0.71/2.91 |
| Robust Mixed H-2/∞ | 0.01/1.50 | 0.01/1.99 | 0.18/5.45 | 0.52/8.35 | 0.68/8.96 |

Table 3 demonstrates the range of uncertainty where our proposed mitigation robust controller remains stable in the presence of uncertainty. The uncertainty, δ, is a complex number to represent the uncertainty in both the magnitude and angle of the control signal. Uncertainty in magnitude represents the possibility of the control input being smaller or larger than the control signal due to the uncertainties discussed in Section 3. Uncertainty in phase represents the uncertainty added by a time delay in the feedback control channel due to communication channel delays. The skewness, σ, is used to study the stability of the controller when the uncertainty is biased in a given direction. This means that we consider the case in which the uncertainties can either increase or decrease the gain of the feedback control input, $\Delta PQ_{EV}$. The actual value of the EVCS real and reactive power load/injections can be smaller than the calculated $\Delta PQ_{EV}$ in the case where attackers compromise some EVCSs or the case where the utility overestimates the number of EVCSs, the control signal is reaching. The actual value of the EVCS real and reactive power load/injections can be larger than the calculated $\Delta PQ_{EV}$ in case the utility underestimates the number of EVCSs the control signal is reaching especially if the clustering mechanism discussed in Section 3 is used.

### 5.2. SMALLER DEFENDER LOAD

Another issue associated with the uncertainty of the feedback EV defender load is that the defender load might become smaller than the attack load. If the attacker successfully compromises enough EVs such that the total defender load is smaller than the total attack load, the defender would be able to partially mitigate the attack impact but not eliminate it. This assumption, however, would require an attacker with huge resources to compromise enough load and EVs as well as keep the EV compromise hidden from the utility. We repeat Attack 2 when the defender controls 25% less load than the attacker and the mitigation scheme successfully reduced the oscillations resulting from the attack to 0.25 Hz (75% reduction).

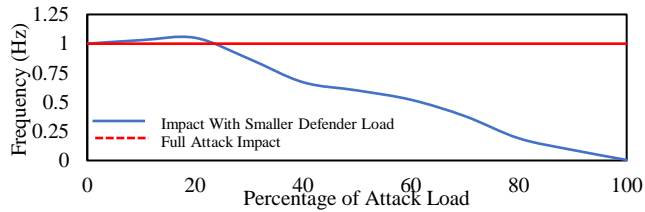

Fig. 26. Maximum frequency deviation vs defender resources level

To demonstrate the impact of limiting the utility's resources below those of the attacker, Fig. 26 presents the maximum frequency deviation achieved under Attack 2 when the utility EV load is capped at different levels. To this end, we use steps of 10% of the total attack load. We can conclude that even as we decrease the defender's resources, the attack impact is still reduced below its original value. It is only after the utility's EV load drops to 20% that the mitigation scheme stops having a positive impact.

Finally, we examine the possibility of an attacker gaining access to a large portion of EVCS (unsecure private EVCSs for instance [60]). However, the modeling of the uncertainty in the feedback load of the controller, $u = \Delta PQ_{EV}$, accounts for such scenario and the mitigation scheme remains successful as long as the attacker does not control a larger portion of the EVCSs than the utility. To put things into perspective, let us assume the attacker attempts to compromise several EVCS in tandem with the previous 800MW attack. Based on the total (public and private) EVCS load curve in Fig. 3 the average EVCS charging load is 2960 MW. Thus, the attacker is required to compromise 1080MW of EVCS load in addition to the original 800MW to start degrading the performance of the mitigation scheme. Under such a condition, the attacker has compromised a total of 1880MW while the defender still has access to 1880MW of EV load. Even attackers knowledgeable of the presence of the mitigation scheme are required to compromise this much EVCS load to have an impact on the performance of our proposed EV-based mitigation. Beyond this point, the aggregate defender load starts becoming smaller than the aggregate attacker load following the behavior in Fig. 26. However, we stress that this scenario is far-fetched and practically impossible. Such an assumption would require an attacker with enough resources to compromise 31% of the total load of the power grid, which is extremely implausible.

### 5.3. EFFECT OF COMMUNICATION DELAY

In this subsection, we consider another aspect usually ignored in similar studies which is the communication and synchronization delay present during real deployment. To this end, we now simulate Attack 3 after adding a random Gaussian delay to each of the feedback signals between 0 and 10ms to examine the impact of the system incurring synchronization delay. The obtained generator frequency response is depicted in Fig. 27. It is evident from this response that the mitigation scheme is still successful despite this delay in the communication channel. To quantify things in terms of controller success, the maximum and average frequency deviations were reduced to 0.026 Hz and 0.013 Hz respectively. This reduction is equivalent to eliminating 98.4% and 98.8% of the maximum and average frequency deviations caused by the LA attack respectively.

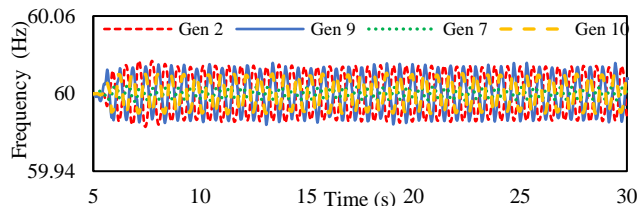

Fig. 27. Generator frequency with controller feedback communication delay

Table 4 also serves to demonstrate the maximum frequency deviation caused by Attack 3 under the robust H-2, H-∞ and mixed sensitivity H-2/∞ control strategy. In this table, the delay is assumed equal in all the feedback channels. Table 4 demonstrates that the EV-based mitigation scheme is successful albeit with slightly reduced performance. This shows that the maximum possible 10ms synchronization delay has been effectively accounted for in the controller synthesis.

TABLE 4. MAX FREQUENCY DEVIATION VS COMMUNICATION/SYNCHRONIZATION DELAY

| Delay (ms) | 0 | 1 | 2 | 4 | 6 | 8 | 10 |
|---|---|---|---|---|---|---|---|
| H-2 | 0.006 | 0.012 | 0.02 | 0.025 | 0.029 | 0.03 | 0.032 |
| H-∞ | 0.1 | 0.13 | 0.17 | 0.19 | 0.21 | 0.24 | 0.26 |
| Mixed H-2/∞ | 0.01 | 0.04 | 0.07 | 0.1 | 0.11 | 0.12 | 0.13 |

### 5.4. IMPACT OF CLUSTERING

In this subsection, we briefly discuss the impact that the clustering suggested in Section 3 would have on the performance of the EV attack mitigation. To this end, we use the example that the utility clusters EVs into groups such that the expected participation is 10 EVs per cluster. To this end, we sample the feedback control signal and send a signal equal to a multiple of the power per cluster. Each cluster is assigned a load value equal to 10 EVs. With this clustering technique, Attack 2 is repeated and results in the frequency responses seen in Fig. 28. It is evident that this technique reduces the effectiveness of the controller, but the achieved response is still within the acceptable range of normal frequency operation. The maximum and average frequency deviations were reduced by factors of 95.4% and 97.4% to reach 0.05 Hz and 0.019 Hz respectively.

## 5.5. IMPACT OF GRID OPERATING POINT

Finally, we examine the impact of the grid's operating point on the success of the LA attack mitigation scheme. To this end, we scale the NE grid based on the NSW load profile [51]. The average load in the NSW grid is 6968 MW while the peak and lowest loads are 8214 and 5897 respectively. We repeated Case Study 4 on the scaled grids and observed that when using the correct state-space matrices, the operating point has very little impact on the performance of our proposed mitigation scheme. This is to demonstrate that even when the dynamic behavior of the power grid changes, the mitigation scheme is still successful as long as the utility maintains updated and correct parameters of their grid in the calculation of the controller gain matrix.

Alternatively, the utility can model the grid's parameter uncertainty by modifying (23) and assigning a non-zero variable to $\Delta A$ in (24). Consequently, the uncertainty in matrix A of the state-space representation can be modeled using steps resembling (26) to (37) that were used to represent the uncertainty in the feedback loop and matrix B. However, as mentioned earlier, it is widely accepted to model the power grid as a state space [37]- [39] and to use this state space for controller design [40]- [42] while assuming the utility has the complete accurate knowledge of their own power grid parameters. Finally, in the farfetched case that a utility does not have an accurate state-space representation, they can follow the approach proposed in [15] to obtain an accurate state-space representation. This approach utilizes a system identification technique consisting of introducing a small probing signal and retrieving the grid's impulse response to create the state-space matrices using the Eigenvalue Realization Algorithm [61].

## 6. CONCLUSION

The exponentially growing number of EVs coupled with their presence on the load buses of the grid and fast communication infrastructure, make them ideal for our EV-based LA attack mitigation scheme. In this work, we demonstrated how EVs can be modeled as a feedback loop controller used to stabilize the grid under LA attacks. The controller synthesis is based on the state-space model of the grid which is needed to design the feedback gain based on robust mixed H-2/∞ control. The mixed controller design incorporates the uncertainty in the feedback EV signal to achieve grid stability under static, switching, and dynamic attacks. The initial 1.5 Hz frequency deviation caused by an 800MW attack was attenuated by our scheme below 0.01 Hz guaranteeing system stability. The controller was also successful under different operating conditions such as EV load uncertainty and communication delay. We also presented an EV clustering technique that can be used to preserve the privacy of home EVCSs. The performance of the controller did vary slightly when clustering was implemented, but the result remained well within the normal frequency behavior of the grid.

### ACKNOWLEDGMENT

The work of Ph.D. candidate M. A. Sayed is supported by Fonds de Recherche du Quebec (FRQNT), project 2022-2023 - B2X - 317973. This research was conducted and funded as part of the Concordia University/ Hydro-Quebec/ NSERC research collaboration project "Large-Scale Integration of EVCSs into the Smart Grid" Grant reference: ALLRP 567144-21.